\documentclass[mathleft
]{an}
\usepackage{graphicx}
\usepackage{times}
\usepackage{natbib}
\bibpunct{(}{)}{;}{a}{}{,}
\begin{document}

\Pagespan{789}{}
\Yearpublication{2006}%
\Yearsubmission{2005}%
\Month{11}%
\Volume{999}%
\Issue{88}%

\title{The Blazhko modulation of TV Boo in 2010}

\author{G. Hajdu\inst{1}\fnmsep\thanks{\email{hajdu@konkoly.hu}\newline}
\and  J. Jurcsik\inst{1}
\and  \'A. S\'odor\inst{1,2}
\and  B. Szeidl\inst{1}
\and  P. Smitola\inst{1}
\and  B. Belucz\inst{3}
\and  K. Posztob\'anyi\inst{4}
\and  K. Vida\inst{1}
\and  E. Kun\inst{5}
}
\titlerunning{The Blazhko modulation of TV Boo}

\authorrunning{G. Hajdu et al.}
\institute{
Konkoly Observatory, MTA CSFK, Konkoly Th. M. \'ut 15-17., 1121 Budapest, Hungary
\and
Royal Observatory of Belgium, Ringlaan 3, B-1180 Brussels, Belgium
\and
E\"otv\"os University, Dept. of Astronomy, H-1518 Budapest PO Box 49, Hungary
\and
Visiting observer at Konkoly Observatory
\and
Department of Experimental Physics and Astronomical Observatory, University of Szeged, H-6720 Szeged, D\'om T\'er 9, Hungary
}

\received{30 May 2005}
\accepted{11 Nov 2005}
\publonline{later}


\abstract{%
  We present the analysis of the multicolour CCD observations in the 2010 season of the Blazhko RRc star TV~Boo. TV Boo shows a complex Blazhko modulation dominated by two independent modulations with $\textrm{P}_{fm1}=9.74$ d and $\textrm{P}_{fm2}=21.43$ d long periods. Both modulation components appear in the frequency spectra as multiplet structures around the harmonics of the pulsation. The positive value of the asymmetry parameter ($Q=+0.51$) of the primary modulation suggests that it is similar in its character to the Blazhko-effect of most of the modulated RRab stars. Interestingly, the secondary, lower-amplitude modulation exhibits a negative asymmetry parameter ($Q=-0.22$), which is an unusually low value when compared to the Blazhko-modulated RRab stars. Apart from the two modulation frequencies, the spectra also show an additional frequency  $f'$ and its linear combinations with the pulsation and the primary modulation $f_{1}+f'$ and $f_{1}-f'-f_{\mathrm{m}1}$. We conclude that the additional frequency most probably belongs to a non-radial mode.}

\maketitle

\section{Introduction}
\label{intro}
More than 100 years after its discovery, the enigmatic Blazhko effect is still one of the greatest challanges to the theory of stellar pulsation. In recent years, both multicolour gound-based observations \citep{kbs1} and the space-based missions $CoRoT$ \citep{corot} and $Kepler$ (\citealt{kepler2}; \citealt{kepler}) greatly contributed to our knowledge of the Blazhko effect in fundamental-mode RR~Lyrae variables (RRab stars).

However, first-overtone (RRc) variables have largely been neglected in these studies due to their smaller number and the noticeably smaller occurrence rate \citep{nagy} of the Blazhko effect among them. As none of the Field-of-Views of space missions contain a single Blazhko-modulated RRc star (J.~Benk\H{o} - private communication), ground-based observations remain the sole possibility of studying these stars.

\citet{olech} reported the presence of double-mode RR Lyrae stars in $\omega$ Cen with a dominant first overtone and period ratios with the secondary frequency of $\sim$0.80 and $\sim$0.61. They conclude that the former ratio can be interpreted as a double-mode RR Lyrae variable with first and second overtone pulsation. However, the second period ratio cannot be explained by radial pulsation. Consequently, the second frequency must belong to a non-radial mode. Additional frequencies have also been reported in a number of RRab stars (see \citealp{xyauzv} and references therein), however, with less consistent period ratios.

Our study aims to analyse the multicolour light-curve behaviour of the birght ($V\sim10.9$ mag), northern, previously known Blazhko-modulated RRc star \citep{detre} TV~Boo ($\alpha_{2000}=14^{\textrm{h}} 16^{\textrm{m}} 36^{\textrm{s}}\!\!.6$ $\delta_{2000}=+42{\degr} 21' 35.\!\!^{\prime\prime}7$).

Recently, \citealp{pena} determined the physical parameters of TV~Boo using both Str\"omgren $uvby\beta$ photometry and the Fourier-decomposition method. However, they did not find indication of the Blazhko effect in their limited observational material, because it is confined mostly to one Blazhko phase. The Blazhko modulation alters the Fourier-parameters of the light curve, consequenty deriving the physical parameters of TV~Boo from the Fourier-decomposition of only one Blazhko phase might be misleading.

\section{Observations}
We observed TV~Boo  between March 2010 and July 2010 on 62 nights with the 60-cm telescope of the Konkoly Observatory at Sv\'abhegy, Budapest, equipped with a 750 $\times$ 1100 pixel Wright Instruments CCD camera, corresponding to 17' $\times$ 27'. Standard Johnson $BVI_{\mathrm{C}}$ filters were used. About 3900 exposures were taken in each band. Relative magnitudes of TV~Boo were estimated against the local comparison star 2MASS~14164052+4224009 using aperture photometry. The magnitudes were transformed to the Johnson-Cousins system utilizing standard procedures.

\begin{figure}
\centering
\includegraphics[width=80mm]{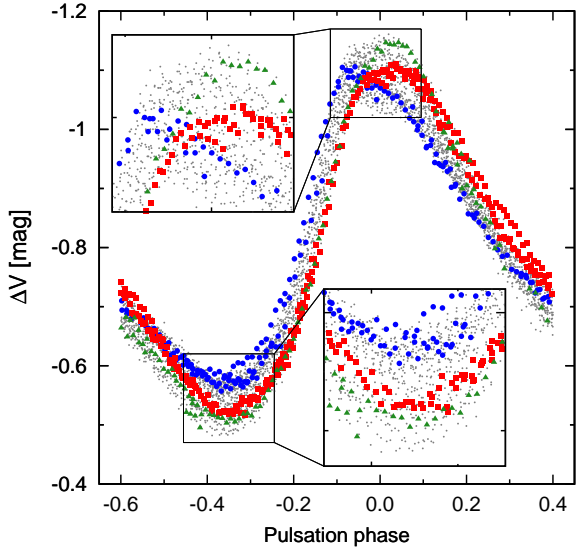}
\caption{Phase-folded $V$ light curve of TV Boo. All observations are plotted by grey points. The squares (red), circles (blue), and triangles (green) correspond to measurements in selected phases of the $\sim9.7$ d primary Blazhko cycle. The inserts magnify the light curve around the maximum and minimum of the pulsation. The existence of amplitude- and phase-modulation is evident. Note that the Blazhko effect varies only the depth of the minima, while the shape and times of the maxima change strongly.}
\label{lc}
\end{figure}

The $V$ light curve of TV Boo folded by the pulsation period is shown in Fig.~\ref{lc}. The amplitude of the pulsation is $\sim$ 0.55 mag, while the Blazhko effect varies the height of the minima and the maxima by $\sim$ 0.1 mag.

\section{Analysis}
We analyse the light curve of TV~Boo with the period-analysis package MUFRAN \citep{mufran}, the GNUPLOT utilities\footnote{http://www.gnuplot.info} and a custom developed nonlinear fitting package \citep{lcfit}. We calculate the Discrete Fourier Transform of the light curve and identify the dominant peaks in the residual spectra. New peaks are added successively to the frequency solution. We accept frequencies as real signals if they have an amplitude larger than $2.5\sigma$ above the mean level of the spectrum in at least two bands.

We check whether TV~Boo displays additional frequencies with frequency ratios of $\sim0.80$ and $\sim0.61$ (similar to $\omega$ Cen RR Lyrae stars, see Sect. \ref{intro}) in its residual spectra, and find no significant peaks in the vicinity of these frequency ratios.

The values of the independent frequencies are derived by fitting the final list of frequencies to the $V$ light curve, while keeping the frequencies of the triplet, quintuplet and linear combination terms locked to the frequency values of the independent ones. Table~\ref{freq} gives the final frequency and corresponding period values of the detected independent frequencies.

\begin{table}
\centering
\caption{Frequencies and periods of the detected independent frequencies.}
\label{freq}
\begin{tabular}{l|cr@{.}l}
                  &Frequency (cd$^{-1})$&\multicolumn{2}{c}{Period (d)}\\
\hline
$f_{1}$           &3.199374             & 0&3125611\\
$f'$              &3.083882             & 0&3242666\\
$f_{\mathrm{m}1}$ &0.102707             & 9&74\\
$f_{\mathrm{m}2}$ &0.046673             &21&43\\
\end{tabular}
\end{table}

The final frequency solution contains \textit{45} frequencies:\\
10 harmonics of the pulsation frequency $kf_{1}$,\\
13 triplet $kf_{1}\pm f_{\mathrm{m}1}$  and\\
5 quintuplet $kf_{1}\pm 2f_{\mathrm{m}1}$ components of the main modulation ($f_{\mathrm{m}1}$),\\
1 the main modulation frequency $f_{\mathrm{m}1}$;\\
13 triplet components $kf_{1}\pm f_{\mathrm{m}2}$ of the secondary modulation ($f_{\mathrm{m}2}$);\\
1 supposed additional frequency $f'$ (see Sect.~\ref{additional}),\\
2 combination frequencies $f_{1}+f'$ and $f_{1}-f'-f_{\mathrm{m}1}$.

\section{Discussion}

Our analysis of the light curve of TV~Boo has revealed that besides normal pulsation, TV~Boo exhibits \textit{two different modulations} and \textit{one additional frequency}: $f_{\mathrm{m}1}$,$f_{\mathrm{m}2}$ and $f'$. Figure~\ref{schematic} shows the schematic plot of the detected frequencies at the low-frequency region and the first five harmonics of the pulsation.

\subsection{Characteristics of the modulations}

\begin{figure}
\centering
\includegraphics[width=80mm]{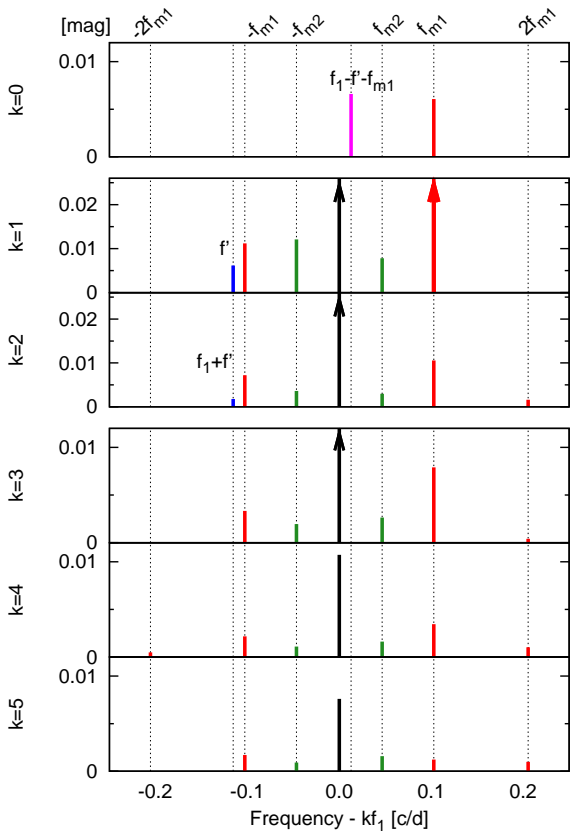}
\caption{Schematic spectrum of the $V$ light-curve solution at the low-frequency region (top panel), and the harmonics of the pulsation frequency $kf_{1}$, $k=1...5$. Detected frequencies are indicated by the vertical lines, with length corresponding to their amplitudes. The color of the lines denote their identification: black, red, green, and blue lines are belonging to the pulsation $f_1$, to the $f_{\mathrm{m}1}$ and $f_{\mathrm{m}2}$ modulations, and to the $f'$ additional frequency, respectively. The combination frequency $f_1-f'-f_{\mathrm{m}1}$ is delineated by its mixed colour. The vertical dotted lines marks the positions of the frequency combinations.}
\label{schematic}
\end{figure}

The primary modulation $f_{\mathrm{m}1}$ has a similar character as the Blazhko-effect of RRab stars, with a triplet structure around most, and quintuplet components around some of the harmonics of the pulsation. The triplet structure around the harmonics of the pulsation are dominated by the higher-frequency ($kf_1+f_{\mathrm{m}1}$) components. The amplitude difference between the two modulation components of the triplet is characterized by the asymmetry parameter: $Q=(A_+ -A_-)/(A_+ + A_-)$, introduced by \citet{alcock} for the triplet structures around the first harmonic of the pulsation. For the primary modulation $f_{\mathrm{m}1}$, $Q=+0.51$, which is a normal value compared to asymmetry values of the LMC \citep{alcock} and the Konkoly Blazhko Survey I and II \citep{kbs2} RRab samples. The modulation frequency itself is clearly present in the spectrum (top panel of Fig.~\ref{schematic}).

The secondary modulation $f_{\mathrm{m}2}$ also shows triplet components, but no quintuplets are found. Neither the high-, nor the low-frequency triplet components are clearly dominant in the different harmonic orders (Fig.~\ref{schematic}, $k=1...5$ panels). The asymmetry parameter for $f_{\mathrm{m}2}$ is $Q=-0.22$, which is in contrast with the typically positive values of the asymmetry parameter for RRab stars. The secondary modulation frequency itself is not detected in the spectrum.

\subsection{The additional frequency and its linear combinations}
\label{additional}

We identify an additional frequency, $f'$ in the Fourier-spectrum of the star. Note that it is very close to the low-frequency component of the triplet structure of the primary modulation ($f_1-f_{\mathrm{m}1}$, $k=1$ panel of Fig.~\ref{schematic}) It is also very close to the pulsation frequency $f_{1}$, ruling out the possibility of it being a radial mode. A simple combination frequency with the pulsation, $f_{1}+f'$ also appears in the light-curve of the star. These two peaks might be interpreted as the dublet components of a tertiary modulation with a frequency $f_{\mathrm{m}3}=f_1-f'$. Supposing an ordinary behaviour of this modulation, we expect to detect a signal at $f_1+f_{\mathrm{m}3}$, forming a triplet structure with $f_1$ and $f_1-f_{\mathrm{m}3}$. The detection limit ($2.5\sigma$) at this frequency is $0.00115$ mag in the $V$ band residual spectrum. However, we do not detect a signal larger than the detection limit at this assumed frequency. Adopting the detection limit as the upper bound on the amplitude of the supposed $f_1+f_{\mathrm{m}3}$ signal, we get an upper bound of $-0.69$ on the asymmetry parameter of the tertiary modulation. We conclude that the additional frequency $f'$ most probably belongs to a non-radial mode, however, a tertiary modulation cannot be ruled out completely.

$f_{1}-f'-f_{\mathrm{m}1}$ is a low-frequency combination term (pink vertical line on the top panel of Fig.~\ref{schematic}). Interestingly, this frequency has an amplitude higher than the primary modulation ($f_{\mathrm{m}1}$).

\section{Conclusions}

We have analysed the multicolour light-curve of TV~Boo from the 2010 observing season. We have found that the light-curve modulation of TV~Boo is intrinsically doubly periodic. Such modulation properties have been reported for a number of RRab stars (see \citealt{rzl} and references therein), but this attribute of the Blazhko-effect is largely unexplored in the case of RRc stars.

Besides the pulsation and the light-curve modulation, TV~Boo shows an additional frequency, close to the first harmonic of the pulsation. Linear combinations of this frequency, the pulsation, and the primary modulation can be found in the low-frequency region of the spectrum and at the second harmonic of the pulsation. We conclude that this frequency probably belongs to a non-radial mode.

We found no evidence for additional frequencies with period ratios reported by \citet{olech} for a number of RR Lyrae stars with dominant first-overtone pulsations in $\omega$ Cen.

\acknowledgements
The financial support of the OTKA grant K-81373 is gratefully acknowledged.

\end{document}